\documentclass[prb,aps,superscriptaddress,showpacs,floatfix,12pt]{revtex4}
\usepackage{amsmath}
\usepackage{amssymb}
\usepackage{color}
\usepackage{graphics}
\usepackage{epsfig}
\usepackage{hyperref}
\begin{document}

%\title{Towards and beyond the maximum speed: active control of pulse Cherenkov and
%tunable Terahertz transition radiation of a Josephson vortex}
%\title{Shear waves of a Josephson vortex: exact solutions and time dilation}
%\title{Shear waves in 2D Josephson junctions: exact solutions and time dilation}
\title{Shape waves in 2D Josephson junctions: exact solutions and time dilation}
%\title{Vortex shape excitations in 2D Josephson junctions: exact solutions and time dilation}
%(Versions: "Three terminal fluxon device for controlling Terahertz radiation using Josephson vortices as malleable waveguides")

\author{D. R. Gulevich }
\affiliation{Physics Department, Loughborough University, Leicestershire, LE11 3TU, United Kingdom}
\affiliation{Advanced Science Institute, The Institute of Physical and Chemical Research (RIKEN), Wako-shi, Saitama, 351-0198, Japan}

\author{F. V. Kusmartsev}
\affiliation{Physics Department, Loughborough University, Leicestershire, LE11 3TU, United Kingdom}

\author{Sergey Savel'ev}
\affiliation{Physics Department, Loughborough University, Leicestershire, LE11 3TU, United Kingdom}
\affiliation{Advanced Science Institute, The Institute of Physical and Chemical Research (RIKEN), Wako-shi, Saitama, 351-0198, Japan}

\author{V. A. Yampol'skii}
\affiliation{Advanced Science Institute, The Institute of Physical and Chemical Research (RIKEN), Wako-shi, Saitama, 351-0198, Japan}
\affiliation{Usikov Institute for Radiophysics and Electronics, Ukrainian Academy of Science, 61085 Kharkov, Ukraine}

\author{Franco Nori}
\affiliation{Advanced Science Institute, The Institute of Physical and Chemical Research (RIKEN), Wako-shi, Saitama, 351-0198, Japan}
%
%Center for Theoretical Physics
\affiliation{MCTP, Department of Physics, University of Michigan, Ann Arbor, MI 48109, USA}
\date{\today}
\begin{abstract}
%Class of exact solutions, keeping shapes, arbitrariness, time dilation -- special relativity. Slowing down of excitations on a %moving vortex.
%relativistic vortices
%to test time dilation -- excitation on a fluxon acts as a ``minute-hand" of the fluxon's inner clock.

We predict a new class of excitations propagating along a Josephson vortex in two-dimensional Josephson junctions. These excitations are associated with the distortion of a Josephson vortex line and have an analogy with shear waves in solid mechanics. Their shapes can have an arbitrary profile, which is retained when propagating. 
We derive a universal analytical expression for the energy of arbitrary shape excitations,
investigate their influence on the dynamics of a vortex line, and discuss conditions where such excitations can be created.
Finally, we show that such excitations play the role of a clock for a relativistically-moving Josephson vortex and suggest an experiment to measure a time dilation effect analogous to that in special relativity. 
%The position of the shape excitation on a Josephson vortex acts like a ``minute-hand" showing the time in the rest-frame associated with the vortex.

%Remarkably, at some conditions, shear wave can carry negative energy: a vortex with the shear excitation can have less energy than the same vortex without it.

\end{abstract}
%05.45.Yv (Solitons)
%03.75.Lm (Tunneling, Josephson effect, Bose–Einstein condensates in periodic potentials, solitons, vortices, and topological excitations)  
%85.25.Cp (Josephson devices), 
%84.40.Az (Waveguides, transmission lines, striplines), 
%84.40.-x (Radiowave and microwave technology)
%85.25.-j (Superconducting devices), 
%74.50.+r (Tunneling phenomena; point contacts, weak links, Josephson effects),
%07.57.Pt (Submillimeter wave, microwave and radiowave spectrometers; magnetic resonance spectrometers, auxiliary equipment, and techniques)

\maketitle

\pagenumbering{arabic}
\pagestyle{plain}

%-----------------------------------------------------------------------------------------------
%-----------------------------------------------------------------------------------------------
%{\bf Introduction.} 
{\it Introduction.---}
%-----------------------------------------------------------------------------------------------
%-----------------------------------------------------------------------------------------------
%Sine-Gordon model describes a variety of physical phenomena including propagation of spin density waves [] (ferromagnets?), information transport in microtubules [], (?) nonlinear optics, the Josephson effect [Josephson, Barone] and penetration of magnetic field high temperature superconductors.
%
The sine-Gordon (SG) model describes a variety of physical systems, including ferromagnets, information transport in  microtubules, dislocations in crystals, nonlinear optics, Josephson junctions and high temperature superconductors %(see,e.g.,~\cite{ferromagnets,microtubules,Bishop,nl-optics,Josephson,Kivshar,Artemenko}). 
(see, e.g.,~\cite{review, Barone, Jackson, Rajaraman}).
The (1+1)D SG model is fully integrable and possesses well known exact solutions in the form of solitons and breathers~\cite{Jackson, Rajaraman}. Depending on the physical realization, they describe different objects, for instance, magnetic domain walls in ferromagnets or vortices (fluxons) in Josephson junctions. There exist mathematical techniques that allow constructing solutions of the (1+1)D SG, such as the B\"{a}cklund transformation, the Hirota and inverse scattering methods (see, e.g., ~\cite{Jackson, Rajaraman}).
Although real systems described by the SG model may contain additional terms associated with damping or external conditions,
such terms can be taken into account as perturbations to the exact solutions~\cite{McS,perturbedSG}.

To describe realistic systems, such as two-dimensional Josephson junctions, one should depart from the well studied (1+1)D model and deal with the (2+1)D SG equation, which is no longer fully integrable. 
%no solution exist in general form. 
The extra dimension opens new avenues for the existence of phenomena which are absent in the 1D case (see, e.g.,~\cite{radial, 2D-effects, cloning, sigma-pump}).
%So far the work with 2+1 SG equation was predominantly numeric.
Here we show that there exists a class of exact solutions of the (2+1)D SG equation which describes the propagation of 
%excitations of an {\it arbitrary} shape 
distortions of an {\it arbitrary} shape along a Josephson vortex line.
%We will show that such solutions can be used to experimentally test time dilation associated with the Lorentz invariance of the sine-Gordon equation which has a direct analogy with the special relativity. 
%Such solutions exhibit the time dilation effect analogous to that in special relativity. 
%associated with the Lorentz invariance of the sine-Gordon equation and are 
%So far, only the Lorentz contraction has been measured in conventional Josephson junctions~\cite{Lorentz-contraction}, but no attention to the time dilation was devoted. 
The property of these excitations such as transmitting pulses of electromagnetic radiation of arbitrary shape along a Josephson vortex can be useful for the transmission of information in various 
% high-frequency 
Josephson devices (see, e.g., theoretical and experimental works~\cite{Saveliev, Koshelets}). 

Experiments~\cite{Lorentz-contraction} have shown that Josephson vortices can exhibit Lorentz contraction. So far, another relativistic effect~\cite{footnote-rel}, time dilation, has been considered difficult to detect in Josephson systems. In order to observe time dilation, an internal degree of freedom acting as a clock is needed. In (1+1)D such clock is absent because solitons of the (1+1)D SG equation do not possess internal oscillation modes. Recently, time dilation was proposed~\cite{time-dilation} in junctions obeying the double sine-Gordon equation. There, an internal degree of freedom of a vortex, due to a special feature of the double SG model was proposed as a clock for a moving vortex. Here we show that there is no need to deviate from the original pure sine-Gordon nonlinearity because
an additional degree of freedom due to an extra dimension, can be used to realize a clock. 
%Thus, a conventional Josephson junction obeying the (2+1)D SG model can be used. 
%
%Apart from the mathematical beauty, 
%of the found solutions they can be possibly used in practice for information transmission and in
%in digital superconducting systems based on conventional Josephson junctions or high temperature superconductors. 
%the solutions describing propagating of electromagnetic waves and excitations in superconducting Josephson junctions are of potential practical importance. Recently a prototype analog-to-digital converter (ADC) was developed by HYPRES that prompts that
%high-speed digital superconducting systems,
%superconducting systems can be of future use for such applications as wireless communication, radars, switching networks and THz technology~\cite{HYPRES}. On the other hand, integrated sub-THz receivers based on long Josephson junctions have been developed recently~\cite{Koshelets}. 

%for information transmission in Terahertz or sub-Terahertz technology and future communications.

%~\cite{Saveliev} 

%-----------------------------------------------------------------------------------------------
%-----------------------------------------------------------------------------------------------
{\it Model.---}
%-----------------------------------------------------------------------------------------------
%-----------------------------------------------------------------------------------------------
%To illustrate the properties of the new class of solutions, 
Consider a 2D Josephson junction described by the (2+1)D SG equation,
\begin{equation}
\varphi_{tt}-\varphi_{xx}-\varphi_{yy}+\sin\varphi=0
\label{2DSG}
\end{equation}
where $\varphi$ is the superconducting phase difference across the Josephson junction. Here, the coordinates $x$ and $y$ are normalized by the Josephson penetration length $\lambda_J$, and the time $t$ is normalized by the inverse Josephson plasma frequency $\omega_p^{-1}$~\cite{Barone}. In the case of a Josephson junction, the soliton solutions of Eq.~\eqref{2DSG} describe Josephson vortices or fluxons~\cite{Barone, Jackson, Rajaraman, McS}.
%A special class of solutions of the (2+1)D SG equation can be constructed 
%%by generalization of the solution of (1+1)D SG which does not depend on one of the coordinates, or, obtained 
%by Lorentz transforming the solutions of a stationary 2D SG equation. 
%%More sophisticated solutions cannot be derived from (1+1)D, and represent essentially two dimensional excitations. An example is radial breathers or pulsons~\cite{radial}. 
%However, there are numerical solutions of the (2+1)D SG equation which cannot be derived from the (1+1)D model, e.g., radial breathers or pulsons~\cite{radial}. 

%-----------------------------------------------------------------------------------------------
%-----------------------------------------------------------------------------------------------
%{\it Shear waves on a Josephson vortex.---}
%-----------------------------------------------------------------------------------------------
%-----------------------------------------------------------------------------------------------
A more general class of solutions of the (2+1)D SG equation has the following form,
\begin{equation}
\varphi(x,y,t)=4\,\arctan\exp\left[y-f(x\pm t)\right]
\label{phi}
\end{equation}
which satisfies Eq.~\eqref{2DSG} exactly with an arbitrary real-valued twice-differentiable function $f(\xi)$. The excitations, described by $f$, are associated with a local shift of a Josephson vortex in the transverse direction and have analogies with elastic shear waves in, e.g., solid mechanics and seismology~\cite{shear-waves}.
Such distortions propagate along the stationary vortex line with the fixed speed $|u|=1$ in units of the Swihart velocity $\bar{c}=\lambda_J\,\omega_p$. 
%As in solid mechanics (e.g.,~\cite{shear-waves}), we can derive the shear modulus for a Josephson vortex (equal to 8 per single vortex). 
Hereafter, we will refer to these excitations
%arbitrary shape excitations 
%described by $f$ 
 as {\it shape waves}.
 
%In the case of a stationary vortex line, the expression for energy of the shape excitation can be derived analytically.
%exactly. 
%Defining the total energy of the excitation as the difference of the energy of the solution Eq.~\eqref{phi} and the energy of %the vortex without excitation ($f=0$), we obtain an expression for the total energy $\Delta E$ as a functional of~$f$: 
%$\Delta E[f]=8\int f'(x)^2\, dx\, >\, 0$.
%\begin{equation}
%\Delta E[f]=8\int f'(x)^2\, dx\, >\, 0
%\label{energy-static}
%\end{equation}
 
\begin{figure}[!tp]
\begin{center}
\includegraphics[width=5in]{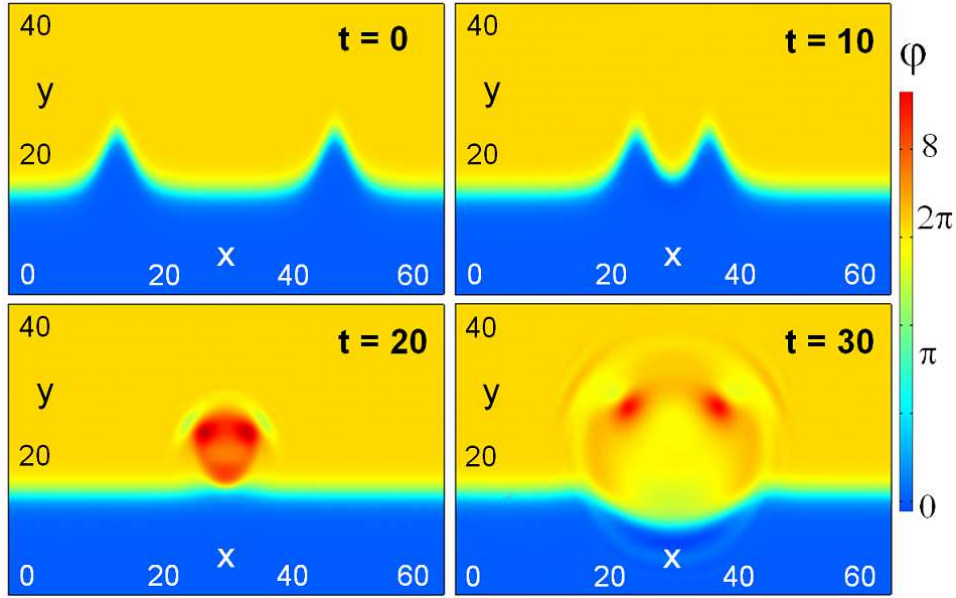}
\caption{\label{collision} (Color online) Collision and annihilation of two large-amplitude solitary shape waves propagating along a Josephson vortex line, Eq.~\eqref{phi} with $f=8/\cosh[(x-x_0\pm t)/2]$. The color scale represents the superconducting phase difference $\varphi$. The figure shows $\varphi(x,y,t)$ snapshots obtained by numerically solving the 
(2+1)D SG equation on a region of size $60\times 40\,\lambda_J^2$ with Neumann boundary conditions: $\mathbf{n} \cdot \nabla \varphi=0$. Initially, the two solitary waves, moving at speed $|u|=1$ toward each other, keep their shape while propagating, but their collision at $t\sim15$ leads to their mutual annihilation and radiation of energy in all directions.
%in the form of plasma waves and two-dimensional breathers. 
Contrary to the large-amplitude waves shown here, small-amplitude shape waves satisfying $|f'(x+t)f'(x-t)|\ll 1$, 
%for all values of $x$ and $t$, 
interact weakly during the collision and behave similarly to solitons. 
}
\end{center}
\end{figure}

%-----------------------------------------------------------------------------------------------
%-----------------------------------------------------------------------------------------------
%{\it Solitary waves.---}
%-----------------------------------------------------------------------------------------------
%-----------------------------------------------------------------------------------------------
%Because of the arbitrariness of $f$, Eq.~\eqref{phi} can describe various physical phenomena.
Because of the arbitrariness of $f$, Eq.~\eqref{phi} describes a variety of excitations of various shapes.
% that will be described below.
Choosing~$f$ localized in a finite area, e.g., $f=A/\cosh(x-t)$, Eq.~\eqref{phi} describes an excitation, 
localized along $x$ that keeps its shape when propagating, i.e., a solitary wave~\cite{Rajaraman}. 
For each solitary wave of this type, there exists an anti-partner with an $f$ of opposite sign in Eq.\eqref{phi}.
 For solitary waves to be solitons, there is an additional important criterion: restoring their shapes after they collide. Consider a trial function
 $ \varphi(x,y,t)=4\,\arctan\exp\left[y-f(x+t)\pm f(x-t)\right]$
that, when $t\rightarrow-\infty$, describes the propagation of two solitary shape waves toward each other (minus sign)
or  a solitary wave and its anti-partner (plus sign). 
One can see that~\eqref{phi} can only {\it approximately} satisfy Eq.~\eqref{2DSG} when $|f'(x+t)f'(x-t)|\ll 1$ for all values of $x$ and $t$. This suggests that, in general, the condition for restoring the shapes may not be satisfied.
%In general case, Eq.~\eqref{2DSG} can not be satisfied, that prompts that the collision of two solitary waves leads to distortion of the original excitations. 
Indeed, our numerical solutions of the (2+1)D SG equation show that the waves keep their shapes when propagating, 
but their collision is destructive.
%but do not tolerate collision with each other. 
%For example, Fig.~\ref{collision} shows that some energy is radiated in the form of plasma waves and breathers, and the shapes become distorted after the collision.
The collision of two waves of large amplitudes may lead to their mutual annihilation, dissipating energy away from the vortex, see Fig.~\ref{collision}. However, smaller-amplitude waves 
%satisfying the above criterion 
 behave similarly to solitons, dissipating very small energy and keeping their original shapes after collision.

Since the equation~\eqref{2DSG} is Lorentz-invariant, one can obtain other solutions performing Lorentz transformations on Eq.~\eqref{phi}. A Lorentz transformation along the $x$-axis
%\begin{equation*}
%\begin{cases}
%x\rightarrow x'=x\\
%y\rightarrow y'=\frac{y-v\, t}{\sqrt{1-v^2}}\\
%t\rightarrow t'=\frac{t-v\, y}{\sqrt{1-v^2}}
%\end{cases}
%\end{equation*}
leads to a Lorentz contraction of the excitation, 
%in the $x$-direction, 
 but does not lead to any solution outside the class of   Eq.~\eqref{phi}, only rescaling the arbitrary function $f(\xi)$. The Lorentz transformation along $y$ turns out to be far more interesting, leading to a class of solutions more general than~Eq.~\eqref{phi}. After such transformation, 
% Eq.~\eqref{phi} becomes
%DOUBLE COLUMN
%\begin{multline}
%\varphi(x,y,t)=\\
%4\,\arctan\exp\left[\frac{y-v\, t}{\sqrt{1-v^2}}-f\left(\frac{t-v\, y}{\sqrt{1-v^2}}\pm x\right)\right]
%\label{phi-Lorentz}
%\end{multline}
%
%SINGLE COLUMN
\begin{equation}
\varphi(x,y,t)=
4\,\arctan\exp\left[\frac{y-v\, t}{\sqrt{1-v^2}}-f\left(x\pm\frac{t-v\, y}{\sqrt{1-v^2}}\right)\right]
\label{phi-Lorentz}
\end{equation}
which describes a Josephson vortex line moving with $-1<v<1$ and a propagating shape excitation described by~$f(\xi)$.
%The $y$-dependence of the function $f(\xi)$ is intriguing: for nonzero $v$ the excitation profile becomes unusually tilted with respect to the vortex line.
%the excitation is no longer perpendicular to the vortex line, but has acquired some finite slope.

%It is known that a Josephson vortex can experience a Lorentz contraction and this has been confirmed by experiments~\cite{Lorentz-contraction}. 
%Here we show that a propagating Josephson vortex can also exhibit time dilation. 

We will show that, apart from the Lorentz contraction~\cite{Lorentz-contraction}, a propagating Josephson vortex exhibits time dilation which influences the dynamics of its shape excitations, Eq.~\eqref{phi-Lorentz}.
Recently, the Lorentz time dilation of a bound half-fluxon pair has been numerically studied~\cite{time-dilation} for a long Josephson junction with a ferromagnetic insulator. The use of the internal degree of freedom of a bound half-fluxon pair was proposed there to test the time dilation.
%detecting the Lorentz reduction in its frequency as a function of pair velocity.
It was found that internal oscillations of the bound pair can play the role of a clock~\cite{time-dilation}.
A Lorentz reduction of their frequency was numerically observed~\cite{time-dilation} when the bound half-fluxon pair was moving with relativistic velocities.
%This internal oscillation acts as a clock.
Here we demonstrate that, to observe the time dilation effect there is no need to use the double sine-Gordon equation~\cite{time-dilation} but the additional degree of freedom of the (2+1)D SG equation can be used to realize the internal clock. Indeed, the {\it shape excitation moving along a fluxon} provides such a clock. In other words, the position of the excitation on a fluxon acts as a ``minute-hand".
From Eq.~\eqref{phi-Lorentz} the position of a vortex changes with time as $y_0(t)=v\,t$,
%, i.e. it moves with the velocity $v$.
%In the ``framework of a vortex".
while the position of the excitation is $x_{0}(t)=\mp [t-v\, y_0(t)]/(1-v^2)^{1/2}=\mp t\,(1-v^2)^{1/2}$.
Thus, the excitation moves along $x$ with speed $|u|=|\dot{x}_0|=(1-v^2)^{1/2}<1$.
In other words, all dynamic processes related to a relativistic moving vortex are {\it slowed down by a time dilation factor $(1-v^2)^{1/2}$, as in special relativity}, compared to the stationary vortex. Our numerical calculations confirm that the collision of two solitary waves is also slowed down, compared to the collision at the rest frame as in Fig.~\ref{collision}.

%The function $f$ can be not localized as considered above, but take non zero values all over the real axis. A special interesting case is a kink propagating along a Josephson vortex, e.g. $f=A\arctan(\exp(t-x))$ or $f=A\tanh(t-x)$. The special case of a kink corresponding to $f=0.75\tanh(t-x)$ was observed in numerical calculations by~\cite{Christiansen}.

%-----------------------------------------------------------------------------------------------
%-----------------------------------------------------------------------------------------------
%{\it Y-dilation---}
%-----------------------------------------------------------------------------------------------
%-----------------------------------------------------------------------------------------------
%The experimental observation of shear excitation can be a challenge  because the short timescales of the order of 100 ps involved, while the shear excitation exists in the inner part of a the Josephson junction, only occasionally reflecting from a boundary. 

%The direct experimental observation of shear excitation can be a challenge because it evolves in the inner part of a the Josephson junction, only occasionally manifestating itself on a boundary.

\begin{figure}[!htp]
\begin{center}
\includegraphics[width=5in]{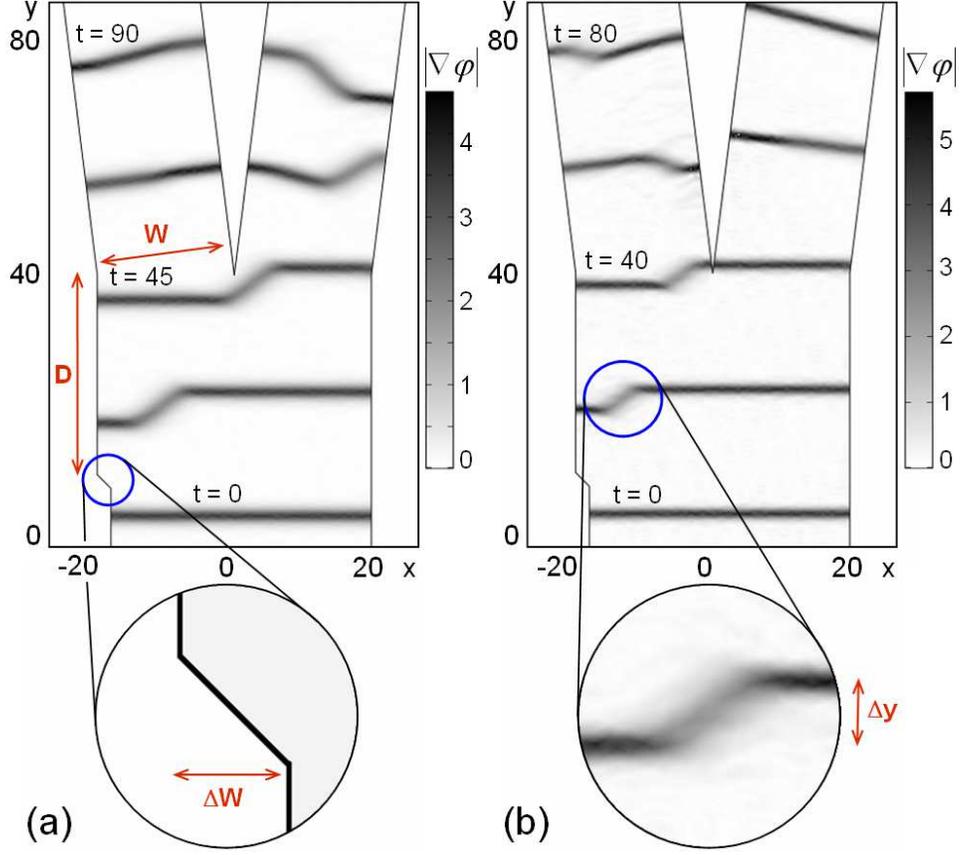}
\caption{\label{Y-dilation} 
%Detection of the shear wave and measurement of the shear wave speed in a system with a Y junction. The shear wave is induced a fluxon line is passing the geometrically induced energy barrier. Then, a propagating fluxon is split by the Y junction in two parts, one containing the shear wave, and the part without the excitation. If a velocity of a vortex is high enough, the time dilation effect affects the apparent propagation of the shear waves seen in the rest frame. As a result, the shear wave in the bottom (left???) branch. The result of the shear wave is to reduce the mean velocity and the total energy of a fluxon, thus 
%affecting the IV characteristic. (a) A vortex propagating with velocity $v=0.8$ and (b) $v=0.9$. 
%In case of (a) the mean velocity of the left part is reduced, and vice versa in (b).
%On (b) the vortex lines appear thinner than (a) because of the Lorentz contraction.
Detection of a vortex shape excitation and measurement of its speed in a Y junction. 
%The scale shows absolute value of the superconducting phase gradient $|\nabla\varphi|$. 
The figure shows snapshots of the superconducting phase gradient $|\nabla\varphi|$ obtained by numerically solving the (2+1)D SG equation 
in a Y junction geometry of size $40\times 80\,\lambda_J^2$ with Neumann boundary conditions: $\mathbf{n} \cdot \nabla \varphi=0$. At time t=0, a fluxon line moving with velocity $v=0.8$~(a) or $v=0.9$~(b) approaches a geometrical energy barrier (shown on the left inset). When passing the barrier, a shape excitation (zoomed on the right inset) is induced on the left end of a fluxon line and propagating to the right along the vortex. The propagating fluxon is split by the Y junction in two parts, one containing the shape excitation, and the part without it. 
Depending on velocity $v$ of the vortex line, the shape excitation can be directed either to the right branch, if the velocity $v$ is low (a), or, to the left branch, if $v$ is higher (b).
%branch of the Y junction 
 This is due to the time dilation reducting the speed of the shape excitation. Measuring the critical velocity $v_{c}$ of the vortex line when such a transition occurs, the speed of a shape excitation can be found as $u=v_{c}\,W/D$.
%The result of the shear wave is to reduce the mean velocity and the total energy of a fluxon, thus affecting the IV characteristic. (a) A vortex propagating with velocity $v=0.8$ and (b) $v=0.9$. 
%In case of (a) the mean velocity of the left part is reduced, and vice versa in (b).
The presence of a shape excitation reduces the average velocity $\langle v \rangle$ of a vortex line, for instance, in (b) the vortex in the left branch at $t=80$ is noticeably delayed from the vortex in the right branch of the Y junction.
%On (b) the vortex lines appear thinner than (a) because of the Lorentz contraction.
}
\end{center}
\end{figure}

%On experiment, the generation of shape waves can be realized in several ways. First, they are excited by inhomogeneities 
%%of a Josephson junction 
% on a way of fluxon movement. 
Experimentally, shape waves can be excited when a fluxon line interacts with inhomogeneities introduced
 on its path. 
%The length of a fluxon line, greater then Josephson penetration length.
%If the length of a fluxon line greater then Josephson penetration length, 
When a vortex line passes through a shaped region of a Josephson junction, 
shape distortions of a vortex line are excited by inhomogeneities of the boundary.
%the interaction with the boundary excites shape distortions the propagate along the vortex line.
%
%results in a delay or acceleration of one end of a fluxon line, creating a shear wave which than propagates to the other end all along the fluxon line. 
%The information about interaction of one end of a fluxon with an inhomogeneity propagates to the other end of a fluxon line in the form of a shear wave. 
%The shape of a shear wave would be defines by the shape of the part of a Josephson junction which a fluxon is passing through. 
The shape of the excited waves is determined by the shape of the inhomogeneity as well as the vortex velocity.
Below we present our analytical and numerical results for the case of a vortex line interacting with a barrier formed by a Josephson junction with a locally-increased width, as shown in Fig.~\ref{Y-dilation}a.
%by the increased width of a Josephson junction. 
Another intriguing possibility to generate the shape excitation
is to use a T junction to create a cloning barrier~\cite{cloning} on the path of a propagating fluxon line.
Other types of inhomogeneities, besides the geometrical ones, such as
microshorts or microresistors, realized by variation of the critical current
%as local variation of the critical current, interactio
 (see, e.g.,~\cite{McS}) can initiate shape excitations.
One more possibility is using injectors of electric current connected locally to a Josephson junction~\cite{Malomed-Ustinov} where the fluxon line is propagating. This allows to conveniently control the shape of the excitations by varying the electric current  of injectors.
%can be used to apply a local magnetic field and locally influence a fluxon line to create a shear excitation. 

%The detection of shear waves is more tricky as the do not manifestate on the IV characteristics, but indirectly trough the influence on the fluxon dynamics.
In order to detect the propagation of a shape wave we propose the following scheme which involves the use of a Y-shaped cloning junction~\cite{cloning}.
Such junction can split a vortex into two parts, which afterward propagate independently from each other. 
For this purpose, a Y junction with a small span angle is preferable, in order to minimize the disturbance on the vortex line. 
When a vortex with a propagating shape wave is split by such a junction, 
%the location of the excitation at the moment of passing the junction can be determined. 
 the shape excitation is directed to one of the branches of the Y junction, see Fig.~\ref{Y-dilation}.
 A shape wave can be generated from a boundary by locally increasing the width of the Josephson junction, as shown in the inset of  Fig.~\ref{Y-dilation}a.
%The alternative methods: localy increased critical current density, cloning barrier, injectors of electric current
Because of the associated energy barrier, one end of the fluxon line is delayed with respect to the other one, which results in a kink profile, as shown in Fig.~\ref{Y-dilation}. The generated kink starts to propagate along the vortex line away from the boundary, carrying information about the induced delay to the rest of the vortex. 

%In the absence of the external forces
%%external current and damping
%and 
%%, provided the boundary barrier is smooth enough, 
%neglecting the radiation losses, one can write the conservation of energy for a single propagating vortex  
%%$$ \frac{8\,W}{\sqrt{1-v^2}}=\frac{8\,(W+\Delta W)}{\sqrt{1-v^2}}+\Delta E $$
%%where $\Delta E$ plays a role of the energy of the shear kink. 
%and find the energy of the shape kink as a difference of the energy of a vortex line before and after the barrier, 
%$\Delta E=-8\,\Delta W/(1-v^2)^{1/2}<0$, where $\Delta W$ is the increase in the vortex length after the barrier, see Fig.~\ref{Y-dilation}. Thus, in contrast to the stationary vortex, 
%Eq.~\eqref{energy-static}, energy of the shape excitation $\Delta E$ of a moving vortex can take {\it negative} values.

Consider now an arbitrary shape excitation with function $f$ satisfying
$f(-W)\to-\Delta f/2$, $f(W)\to\Delta f/2$ for $W\to\infty$.
The total energy of this shape excitation is the difference of the energy $E[\varphi]$ of the solution Eq.~\eqref{phi-Lorentz} and the energy $E[\varphi]_{f\equiv0}$ of the vortex without excitation:
$\Delta E= E[\varphi]- E[\varphi]_{f\equiv0}$.
% that is a functional of~$f(\xi)$.
%
We obtain the following analytic expression for the total energy:
\begin{equation}
\Delta E=\frac{8}{\sqrt{1-v^2}} \left[ \int_{-\infty}^{\infty} f'(x)^2 dx\,-\, v\,\Delta f\right] ,
\label{DeltaE}
\end{equation}
% we obtain a universal expression
which is exact in the limit $W\to\infty$ and valid for an arbitrary excitation $f(\xi)$ of height $\Delta f$.
As seen from Eq.~\eqref{DeltaE}, the energy $\Delta E$ can take {\it negative} values if $f$ is a kink ($\Delta f\ne 0$)  propagating along a moving vortex line.
% a vortex with a shape kink may have smaller energy than the original vortex of the same length. 
Indeed, this is in agreement with our numerical simulations on Fig.~\ref{Y-dilation} which indicate that the increased width $\Delta W$ of the Josephson junction is compensated by the energy gained due to the propagating kink with negative $\Delta E$. Neglecting radiation losses, we now write the conservation of energy relation for a moving vortex line: $\Delta E+8\,\Delta W\,(1-v^2)^{-1/2}=0$.
Using Eq.~\eqref{DeltaE}, the minimal size of the kink can be estimated:
$\Delta f\ge \Delta f_{\rm min}(\Delta W,v)=\Delta W/v$.
%where $\Delta f_{min}(W,v)$ is defined only by the barrier height $\Delta W$ and velocity of the vortex line~$v$. 
%This expression can be used to estimate the minimal height of the kink when the exact shape of the barrier is not known.
Note that the apparent height $\Delta y$ of a kink propagating along a moving vortex line is Lorentz contracted: $\Delta y=\Delta f\,\sqrt{1-v^2}$.
To describe the kinks in Fig.~\ref{Y-dilation} we use the approximation $f(\xi)=\frac12 \Delta f\tanh\left(2v\,\xi/\Delta f\right)$. Substituting it in~Eq.~\eqref{DeltaE} we obtain
$\Delta E=-\,8\,\Delta f\,v\,(1-v^2)^{-1/2}/3$.
%$$\Delta E=-\,\frac{8\,\Delta f\,v}{3\,\sqrt{1-v^2}}$$
From energy conservation, we calculate the apparent height of the kink to be $\Delta y=3\,\sqrt{1-v^2}\,\Delta W/v$. 
Substituting the value $\Delta W=2$, 
%as in Fig.~\ref{Y-dilation}
 we obtain the following results for the apparent heights of the kinks:
$\Delta y\simeq 4.5$ for $v=0.8$ and $\Delta y\simeq 2.9 $ for $v=0.9$, which agree with the numerical results shown in Figs.~\ref{Y-dilation}a and~\ref{Y-dilation}b, respectively.

% where $\Delta W$ is the increase in the vortex length after the barrier. 
%From~\eqref{DeltaE} we can estimate the minimal height of the kink

%for the propagating vortex line:
%$$ \frac{8\,W}{\sqrt{1-v^2}}=\frac{8\,(W+\Delta W)}{\sqrt{1-v^2}}+\Delta E $$

%where $\Delta E$ plays a role of the energy of the shear kink. 
%and find the energy of the shape kink as a difference of the energy of a vortex line before and after the barrier, 
%$\Delta E=-8\,\Delta W/(1-v^2)^{1/2}<0$, where $\Delta W$ is the increase in the vortex length after the barrier, see Fig.~\ref{Y-dilation}. Thus, in contrast to the stationary vortex,...
%
%In other words, propagation of such a shear excitation slows down the average propagation speed of a  vortex,
%
%An experimentally realistic case is interaction of a long fluxon line with an inhomogeneuty localized on a boundary of a Josephson junction. Consider a model by an increased width. The energy barrier slows down one end of a fluxon line creating an excitation of a kink profile. 

%We model $f$ in Eq.~\eqref{phi-Lorentz} as a kink with a linear slope~$g$,
%\begin{equation}
%f(\xi)=
%\begin{cases}
%0, \quad \xi\le 0\\
%g\,\xi, \quad 0\le \xi \le L \\
%g\, L, \quad \xi\ge L \\
%\end{cases}
%\label{shape-kink}
%\end{equation}

After passing the Y junction, each half of the initial vortex line is directed to a Josephson junction of width~$W$.
The presence of the shape kink slows down the average speed of a vortex line,
% compared to a vortex line without such excitation,
%Define the average velocity of a vortex line moving in a Josephson junction of width $2W$ as
\begin{equation}
\langle v \rangle=-\,\frac{1}{2\pi W}\int\!\!\int \dot{\varphi}\,dx\,dy
%=v-\frac{\Delta f\,(1-v^2)}{W}
=v-\frac{\Delta y\,\sqrt{1-v^2}}{W}
\label{v-average}
\end{equation}
%
%Propagation of such a shape excitation along a vortex of length $W$ slows down the average speed of a vortex line as a whole,
%$$ u=v\left[1-\frac{\Delta X}{W}\sqrt{1-v^2}\right] $$
%where $\Delta X$ being the size of the shear wave in $x$ direction.
%$$ \langle v \rangle \le v\left[1-\frac{g_c\,L_{min}}{v}\,\frac{\sqrt{1-v^2}}{W}\right]  $$
%$$ \langle u \rangle \le v\left[1-\frac{\Delta W\,\sqrt{1-v^2}}{W\,(1-\sqrt{1-v^2})}\right]  $$
%The difference in speed of two part of a vortex can be detected.
%
%As the voltage is proportional to $\langle v \rangle$, there is a step corresponding to a shape wave changing the branch of a Y junction exhibited in the I-V characteristics.
Thus, the presence of a shape wave can be directly detected by measuring voltage on the Josephson junction,
 proportional to $\langle v \rangle$.
%As the voltage is proportional to $\langle v \rangle$, the presence of a shape wave can be directly detected by measuring voltage on the Josephson junction.
A convenient system 
%to measure I-V characteristics 
to study shape excitations would be a Y junction introduced into a ring, forming a $\sigma$-shaped Josephson junction which has been recently realized experimentally with high-$T_c$ %for other purposes~\cite{sigma-pump}. 
% to investigate generation of Terahertz waves~\cite{sigma-pump}.
 in the context of generation of Terahertz waves~\cite{sigma-pump}.

%The systems where dynamics of fluxons through the T and Y junctions is studied, have been recently realized experimentally~\cite{sigma-pump}.

The effect of time dilation is to slow down the shape excitation when the vortex line reaches relativistic velocities. 
If the velocity of the shape excitation is smaller than some critical value, it will be directed to the left branch of the Y junction, as on the Fig.~\ref{Y-dilation}b, while the vortex propagating in the right branch will remain undisturbed. 
On the other hand, if the shape excitation propagates faster, it will be directed to the right branch, Fig.~\ref{Y-dilation}a.
Because of the change in the average velocity of a vortex line Eq.~\eqref{v-average}, a voltage step, corresponding to a shape wave changing the branch of the Y junction, will be exhibited on the $I$-$V$ characteristics.
Finding the velocity of a fluxon line $v_c$ from the position of the voltage step, speed of the shape excitation can be calculated as $u=v_{c}\,W/D$, where $D$ is a distance between the barrier and the Y junction, see Fig.~\ref{Y-dilation}. 
%Measuring the $I-V$ characteristics, the speed of the shape excitation can be calculated as $u=v_{c}\,W/D$, with $v_c$ being the critical velocity of a fluxon line when the shape wave changes the branch of the Y junction, Fig.~\ref{Y-dilation},
%where $D$ is a distance between the barrier and the Y junction. 
In this way, the relativistic factor $(1-v^2)^{1/2}$ for the delay of the shape kink propagation can be tested.
%$$
%\frac{W}{D}\,v_{c}\simeq(1-v_{c}^2)^{1/2}
%$$
%Speed of the shape excitation at different values of a vortex $v_c$ can be probed putting several inhomogeneuties in a row.
%, or, by using a multiple splitting (cloning) junction with more than two branches.

In an underdamped Josephson junction, a fluxon can be accelerated to velocities close to the Swihart velocity, typically of the order $10^7\,{\rm m/s^2}$ for Nb Josephson junctions. The velocity of a fluxon depends both on damping and an applied current~\cite{McS}. 
%for $\alpha=0.01$ and current $\gamma=0.1$ we get $v=0.992$ that far enough to observe the relativistic effects.
For the damping parameter $\alpha=0.01$ and current $\gamma=0.03$ in normalized units~\cite{McS}, it is possible to accelerate a vortex to the velocity $v\simeq0.92$, so that the relativistic time dilation factor would be $(1-v^2)^{1/2}\simeq0.4$.
%The modern technology allows fabrication of 
For instance, for typical ${\rm Nb/Al\!-\!AlO_x/Nb}$ Josephson junctions 
%with parameters of the critical current density $j_c=$ and 
with the plasma frequency $\omega_p/2\pi\sim 50\,{\rm  GHz}$, the propagation time of the shape excitation between the two ends of a vortex line in a configuration as in Fig.~\ref{Y-dilation} would be approximately
0.4~ns for a stationary vortex, while for a vortex moving with velocity $v=0.92$ this value would be relativistically
delayed by 0.6~ns.

%and 1~ns for a vortex moving with velocity $0.92$ Swihart velocities.

%the delay between the two parts of a fluxon

%For Josephson junctions ${\rm Nb/Al\!-\!AlO_x/Nb}$ with critical current densities down to few ${\rm A/cm^2}$ and 
%%that corresponds to the Josephson 
%plasma frequency as low as $\omega_p/2\pi\simeq 10\,{\rm  GHz}$~\cite{low-jc}, the delay between the two parts of a fluxon is of the order of 1 nanosecond that can be detected by an oscilloscope, while the typical propagation time of a shear wave, $W/(1-v^2)^{1/2}$ is delayed by the relativistic factor $(1-v^2)^{1/2}$ and, thus, can be as much as hundreds of nanoseconds. 

%In fact the method of resonance method described Ref. 11 can be also applied to our case. Although, we have described the method which can be done only measuring of the IV characteristics. Moreover, the shear wave excitation represent a new unstudied unique phenomena.

%-----------------------------------------------------------------------------------------------
%-----------------------------------------------------------------------------------------------
%{\bf Conclusions.}
{\it Conclusion.---}
%-----------------------------------------------------------------------------------------------
%-----------------------------------------------------------------------------------------------
%
%The class of solutions Eq.~\eqref{phi-Lorentz} are of crucial importance for description of interaction of moving Josephson vortices with pinning centers inside the Josephson junction or in the boundary. According our calculations, stationary any perturbation of a vortex line decays into two perturbations of the type~\eqref{phi-Lorentz} running in the opposite directions and plasma. 
%
We have predicted the existence of excitations of arbitrary shapes propagating along stationary or moving Josephson vortices in 2D Josephson junctions, and presented the exact solutions of the (2+1)D SG equation. 
Remarkably, at some conditions, a shape wave can carry negative energy, i.e., a vortex with a shape excitation can have smaller energy than the same vortex 
%propagating with the same velocity 
without it.
Based on our analysis, we suggest how to test a time dilation effect analogous to that in special relativity: using the shape excitation as a ``minute-hand" measuring the proper time of the vortex's coordinate frame.
%of the coordinate frame associated with the vortex.

%-----------------------------------------------------------------------------------------------
%-----------------------------------------------------------------------------------------------
%{\bf Acknowledgments.}
%{\it Acknowledgments.---}
%-----------------------------------------------------------------------------------------------
%-----------------------------------------------------------------------------------------------
We acknowledge partial support from the NSA, LPS, ARO, NSF Grant No.~EIA-0130383,
JSPS-RFBR No.~06-02-91200, JSPC-CTC program and MEXT Grant-in-Aid
No. 18740224. Also, support from EPSRC via No. EP/D072581/1,
EP/E042589/1, EP/F005482/1, and ESF network programme ``AQDJJ''.


\begin{references}
\bibitem{footnote-rel} 
%Note that in the relativistic dynamics of Josephson junctions the role of the speed of light is played by the Swihart velocity.
Note that the relativistic dynamics of Josephson junctions is defined by the Swihart velocity $\bar{c}$ which plays a similar role to the speed of light in special relativity.
\end{references}
\end{document}